\documentclass[10pt,letterpaper,twocolumn]{article} 
\usepackage{ol2}
\usepackage[draft]{hyperref}
\usepackage{amsmath}

\begin{document}

\twocolumn[ 

\title{Theoretical Analysis of the Characteristic Impedance in Metal-Insulator-Metal Plasmonic Transmission Lines}

\author{Hamid Nejati$^{1*}$ and Ahmad Beirami$^{2}$}

\address{
$^1$ Electrical Engineering and Computer Science Department, University of Michigan \\ 1301 Beal Ave., Ann Arbor, MI 48105, USA
\\
$^2$ School of Electrical and Computer Engineering, Georgia Institute of Technology\\ 777 Atlantic Dr. NW, Atlanta, GA 30332, USA\\
$^*$Corresponding author: {hnejati@umich.edu}
}

\begin{abstract}We propose a closed form formulation for the impedance of the Metal-Insulator-Metal (MIM) plasmonic transmission lines by solving the Maxwell's equations.
 We provide approximations for thin and thick insulator layers sandwiched between metallic layers. In the case of very thin dielectric layer, the surface waves on both interfaces are strongly coupled resulting in an almost linear dependence of the impedance of the plasmonic transmission line on the thickness of the insulator layer.
 On the other hand, for very thick insulator layer, the impedance does not vary with the insulator layer thickness due to the weak-coupling/decoupling of the surface waves on each metal-insulator interface.
We demonstrate the effectiveness of our proposed formulation using two test scenarios, namely, almost zero reflection in Tee-junction and reflection from line discontinuity in the design of Bragg reflectors, where we compare our formulation against previously published results.
\end{abstract}


 ] 

Surface Plasmon Polariton (SPP) is a TM polarized electromagnetic wave traveling along metal-dielectric interfaces. It can also be explained as an oscillatory motion of electrons bounded to the metal-dielectric interface, where the corresponding field profiles decay exponentially into the neighboring media from their maxima occurring at the interface~\cite{dionne1}. SPP can confine light to the lateral dimension in the order of tenth of the wavelength overcoming diffraction limit for travelling waves in the Metal-Insulator-Metal (MIM) transmission lines~\cite{dionne1,veronis}. MIM lines have been recently used to build microwave-inspired plasmonic devices, such as filters, branch-line couplers, and resonators~\cite{Hamid2,Hamid4}, as well as optical-inspired devices, such as Bragg reflectors, plasmonic photonic crystals, and Mach-Zehnder couplers~\cite{ditlbacher,han,Hamid3}.
An effective design of the aforementioned devices requires a reliable and accurate method to derive the impedance of the MIM line. To the best of the authors' knowledge, there are only a couple of methods to characterize the impedance of the MIM line in the literature (cf.~\cite{Hamid,veronis2005}). However, those methods do not accurately explain the characteristic impedance of the MIM structure for generally designed MIM lines of various thicknesses. The characterization of the impedance of the MIM can be useful for a better understanding of the MIM lines. It also provides a design tool to generate different microwave/optical-based plasmonic structures.

\begin{figure}[t]
\centering
\includegraphics[height=2in, angle=0]{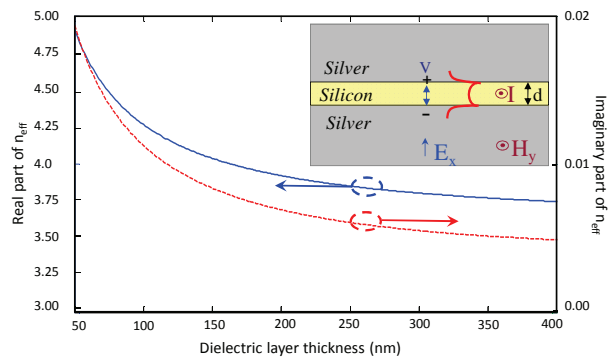}
\vspace{-0.15in}
\caption{\small{Real (blue solid line) and imaginary (red dashed line) parts of the effective refractive index of an MIM structure (shown in the inset) as a function of the dielectric layer thickness. The operating wavelength is assumed to be $1.55\mu m$.}}
\vspace{-0.2in}
\label{fig:neff}
\end{figure}

In the following, we propose a formulation for the impedance, which is derived by solving the Maxwell's equations. This formulation is valid for any thickness of the dielectric layer ranging from very thin to very thick layers. Then, two design scenarios are discussed. A design of a Tee-junction for minimum reflection using the proposed formulation and its comparison to other published formulation methods is presented. The second scenario analyzes the minimum reflection from a line discontinuity using the proposed formulation. This line discontinuity can be utilized in the design of Bragg reflectors.

Starting with plane-wave excitation of the MIM structure and solving the Maxwell's equations, the only propagating mode is the transverse magnetic (TM) mode. The effective refractive index can also be calculated using the solution of the Maxwell's equations~\cite{dionne1}. The effective refractive index for a silicon layer sandwiched between silver layers is shown in Fig.~\ref{fig:neff}. The real and imaginary parts are shown in two different vertical axes and the data are shown as a function of the silicon layer thickness. The effective refractive index will be used in the derivation of the impedance of the transmission lines.

\begin{figure}
\centering
\includegraphics[width=2.7in, angle=0]{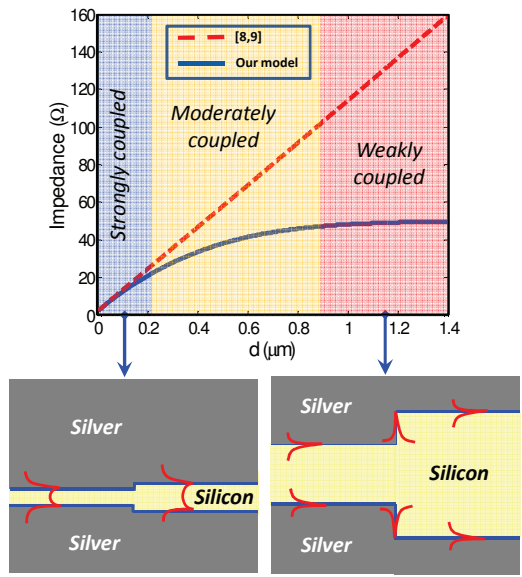}
\caption{\small{MIM characteristic impedance as a function of the thickness of the silicon layer for $W=1\mu m$. The leftmost shaded region is the strongly coupled region, where the impedance is directly proportional to the thickness of the dielectric layer. The middle and right shaded regions are the moderately coupled and the weakly coupled regions, respectively. In the weakly coupled region, the impedance is almost unvarying with the dielectric thickness and the modes are almost decoupled. The sketch of the strongly and weakly coupled regions are also shown in the bottom.}}
\vspace{-0.1in}
\label{fig:impedance}
\end{figure}

Wave-guiding impedance characterizes the wave propagation in the medium. Higher impedance corresponds to harder movement of current carriers and lower impedance corresponds to easier movement of current carriers. The wave impedance is given by
\begin{eqnarray}
Z &=& \frac{V}{I}=\frac{\int_{-\frac{d}{2}}^{\frac{d}{2}}E_Z dz}{\oint_C H_Y dl},
\label{eq:impedance_def}
\end{eqnarray}
where $Z$ is the impedance, $V$ is the voltage difference between metal-insulator interfaces, and $I$ is the travelling current along the line. $C$ is a closed contour inside the dielectric layer and the metal/dielectric interfaces. The electric and magnetic fields ($E_Z$ and $H_Y$), which are given by the solutions to the Maxwell's equations, consist of two terms due to the superposition of the surface waves on each metal-insulator interface. Substituting the resulting fields in~(\ref{eq:impedance_def}) leads to
\begin{eqnarray}
Z &=& \frac{\int_{-\frac{d}{2}}^{\frac{d}{2}} \frac{k_x}{k_{zd}}(e^{-jk_{zd}z}+e^{jk_{zd}z}) dz}{\oint_C \frac{-\omega \epsilon_d}{ck_x}(e^{-jk_{zd}z}-e^{jk_{zd}z}) dl} \nonumber \\
 &=& 2\frac{k_x}{jWk_{zd}\omega \epsilon_d}\tanh{\left(\frac{jk_{zd}d}{2}\right)},
\label{eq:impedance}
\end{eqnarray}
where $\omega$ is the operating radial frequency, $W$ is the unit width of the structure. $k_x$ is the longitudinal wave vector of the propagating surface wave, and $k_{zd}$ is the transversed wave vector of the evanescent wave in the transverse plane with respect to the incident wave. For very thick dielectric layers (large $d$) or high refractive index material (large $n$), the impedance is given by
 \begin{equation}
Z_{\text{thick}} \approx \frac{2k_x}{ W |k_{zd}| \omega \epsilon_d}.
\label{eq:Z_thick}
 \end{equation}
On the other hand, for very small dielectric layer thicknesses (small $d$), the impedance can be approximated by the first dominant term in its Taylor series, i.e.,
 \begin{equation}
 Z_{\text{thin}} \approx \frac{k_x d}{W \omega \epsilon_d}.
 \label{eq:Z_thin}
 \end{equation}
 The resulting term in~(\ref{eq:Z_thin}) is exactly the approximated formula derived previously in~\cite{veronis2005,Hamid}. However, this approximation by the first dominant term is only valid for very thin dielectric thicknesses with low refractive index profiles. The impedance formulation as well as its first dominant term are shown in Fig.~\ref{fig:impedance}. The behavior of the impedance as a function of dielectric thickness is highly linear for very low thicknesses, where the modes are strongly coupled. A hand drawn sketch shows the behavior of the field profiles inside the dielectric layer in Fig.~\ref{fig:impedance}. Increasing the thickness of the dielectric layer leads to an almost constant behavior of the impedance. This phenomenon is due to the decoupling of the modes traveling on each Metal-Insulator (MI) interface. A hand drawn sketch shows the behavior of almost decoupled surface waves on MI interfaces. Fig.~\ref{fig:impedance} also compares the proposed formulation with the other formulations presented in the literature~\cite{veronis2005,Hamid}.
 Next, we present two case studies for the validation of our results.

\begin{figure}[t]
\centering
\includegraphics[width=2.71in, angle=0]{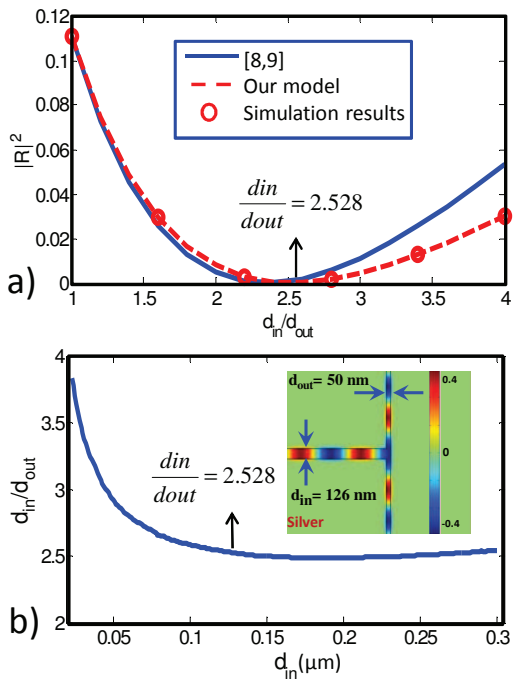}
\vspace{-0.15in}
\caption{\small{(a) The reflection coefficient as a function of the ratio of the input to the output branch thicknesses using models presented in the \cite{veronis2005,Hamid} (blue solid line), our proposed formulation (red dashed line) and the COMSOL results (red circles). (b) The optimized ratio as a function of the input branch thickness. $H_z$ in a Tee-junction is shown in the inset.}}
\vspace{-0.15in}
\label{fig:reflection}
\end{figure}


\textbf{Tee junction design/test scenario:} Tee junction is a power divider (shown in the inset of Fig.~\ref{fig:reflection}(b)), in which the design objective is to minimize the power that is reflected from the junction. The transmitted power on both branches of the Tee junction are equal in magnitude. In order to obtain zero reflection from the Tee-junction, the thicknesses of the dielectric layers should be chosen such that the input impedance matches the impedance of the output branches.

By assuming a fixed thickness of $50nm$ for the output branches, the thickness of the input layer can be calculated using the proposed impedance formulation in~(\ref{eq:impedance}). In Fig.~\ref{fig:reflection}(a), the reflection coefficient is shown as a function of the ratio of the thicknesses of the input and the output layers. Three different methods have been used to derive reflection. The result from the impedance formulation in~\cite{veronis2005,Hamid} (i.e., $Z_{\text{thin}}$) is denoted by the blue solid line in Fig~\ref{fig:reflection}(a). As can be seen, our formulation, shown by the red dashed line, closely follows the full wave simulation (COMSOL) results, shown by the red circles. Although the formulations in~\cite{veronis2005,Hamid} are relatively accurate for very thin dielectric layer thicknesses, the accuracy of the formulations deteriorates as the thickness of the dielectric layer increases.

In Fig.~\ref{fig:reflection}(b), the optimal ratio of the input to the output thicknesses (i.e., where the input and output impedances are matched) is demonstrated as a function of the input branch thickness. By using the optimal ratio, the Tee-junction can be designed for the minimum reflected power. The power flow is shown in the inset of Fig.~\ref{fig:reflection}(b) for input line thickness of $126nm$ and output line thickness of $50nm$.

\textbf{Impedance discontinuity design/test scenario:} Impedance discontinuity is a change in the impedance of the line that is obtained by varying the physical size of the line and/or changing the dielectric material.  The impedance discontinuity can be utilized in the design of Bragg reflectors. In this part, we consider the physical thickness modulation of the line in addition to the dielectric material mismatch on both sides of line discontinuity. Impedance discontinuity results in wave reflection right at the discontinuity.
 The waveguide discontinuity of the MIM lines is studied in~\cite{fan09}, where the modal-analysis of the waveguide modes is used to analyze this structure using the impedance formulation in~\cite{veronis2005}. In the case where different dielectric layers are used at the sides of the discontinuity, an accurate formulation for the impedance is crucial for accurate calculation of the reflection coefficient. Based on the proposed formulation, we consider a thin MIM structure with silica as its insulator, which is connected to a thick MIM structure with silicon as its insulator. In this scenario a thin layer of silica (e.g. $88nm$ thick) is terminated to a thicker layer of silicon (e.g. $300nm$ thick) both sandwiched between silver layers. The respective impedances of the Silicon and Silica layers as well as the reflection coefficients are listed in Table~\ref{tab:comparison}. The impedance values reported by COMSOL are derived using the postprocessing module and integrating over the desired contours. Increasing the number of mesh elements enhances the accuracy between our proposed formulation and the COMSOL results. For the ranges of interest, the previously reported results are very inaccurate and unreliable while the formulation proposed in this letter can be used to accurately obtain the impedance and the reflection coefficients.

\begin{table}[t]
  \centering
  \caption{Reflection coefficient of an impedance discontinuity in MIM structure}
\begin{tabular}{cccc} \\ \hline
    Method & $Z_{\text{Silica}}$ & $Z_{\text{Silicon}}$ & $|R|^2$  \\ \hline
    \cite{veronis2005,Hamid} & 28.15 & 35.49 & 1.33e-2   \\
    Our work & 27.84 & 27.83 & 32e-9  \\
    COMSOL & 27.85 & 28.01 & 8.2e-6  \\ \hline
  \end{tabular}
\vspace{-0.1in}
\label{tab:comparison}
\end{table}

In conclusion, this letter presented a novel formulation for the characteristic impedance of the MIM transmission lines. The proposed impedance formulation is derived by solving Maxwell's equations. This formulation is validated using finite-element simulation software (COMSOL) for two test scenarios including: almost zero reflection from Tee-junction and the impedance discontinuity. The behavior of the impedance formulation was shown to be almost linear as a function of dielectric core thickness for very low thickness and low refractive index materials. On the other hand, the linear behavior approaches a nearly constant impedance for high thickness or high refractive index materials. The almost linear characteristic is due to the strongly coupled modes while the nearly constant behavior of the impedance is due to the weakly coupled modes. We stress that our results are also applicable to moderately coupled region that can not be explained using prior work.

This work was done in part when the authors were affiliated with Rice University. The authors are also grateful to professor Yehia Massoud for bringing the importance of this problem to their attention.

\pagebreak

\section*{Informational Fourth Page}

\end{document}